\documentclass[conference,a4paper]{IEEEtran}
\IEEEoverridecommandlockouts
\usepackage{cite}
\usepackage{amsmath,amssymb,amsfonts}
\usepackage{graphicx}
\usepackage{textcomp}
\usepackage{xcolor}
\usepackage{url}
\usepackage{algorithm}
\usepackage[noend]{algpseudocode}
\usepackage{longtable}
\usepackage{booktabs}
\usepackage{subcaption}

\def\BibTeX{{\rm B\kern-.05em{\sc i\kern-.025em b}\kern-.08em
    T\kern-.1667em\lower.7ex\hbox{E}\kern-.125emX}}
\begin{document}

\title{Trustworthy Intrusion Detection: Confidence Estimation using Latent Space\\}


\author{
    \IEEEauthorblockN{
        Ioannis Pitsiorlas\IEEEauthorrefmark{1}, George Arvanitakis\IEEEauthorrefmark{3}, Marios Kountouris\IEEEauthorrefmark{1}\IEEEauthorrefmark{2}
    }
    \IEEEauthorblockA{\IEEEauthorrefmark{1}\textit{Communication Systems Department, EURECOM, France}}
    \IEEEauthorblockA{\IEEEauthorrefmark{3}\textit{Technology Innovation Institute, Abu Dhabi, UAE}}
    \IEEEauthorblockA{\IEEEauthorrefmark{2}\textit{DaSCI, Department of Computer Science and Artificial Intelligence, University of Granada, Spain}}
}

\maketitle

\begin{abstract}
This work introduces a novel method for enhancing confidence in anomaly detection in Intrusion Detection Systems (IDS) through the use of a Variational Autoencoder (VAE) architecture. By developing a confidence metric derived from latent space representations, we aim to improve the reliability of IDS predictions against cyberattacks. Applied to the NSL-KDD dataset, our approach focuses on binary classification tasks to effectively distinguish between normal and malicious network activities. The methodology demonstrates a significant enhancement in anomaly detection, evidenced by a notable correlation of \textbf{0.45} between the reconstruction error and the proposed metric. Our findings highlight the potential of employing VAEs for more accurate and trustworthy anomaly detection in network security.
\end{abstract}


\section{Introduction}
In the era of digital advancement, the Internet's rapid expansion has been paralleled by a significant increase in sophisticated cyberattacks and as a result, network security has become an important domain \cite{https://doi.org/10.1002/ett.4150}. These attacks not only threaten individual privacy and security but also challenge the integrity of critical infrastructure \cite{electronics12061333}. Against this backdrop, Intrusion Detection Systems (IDS) have emerged as an essential tool in the cybersecurity arsenal, designed to detect and mitigate malicious activities in network traffic \cite{inproceedingsnetwork}. An IDS identifies unauthorized or harmful attacks that frequently occur in a network\cite{8403759}. To address vulnerable attacks, many tools and mechanisms have been developed over the years, with a significant focus on learning-aided algorithms \cite{9483916}\cite{Binbusayyis2022}.

More recently, machine learning (ML) has taken center stage, to make a categorization of different types of network attacks \cite{SOWMYA2023100827}, employing techniques ranging from supervised learning, where models are trained on labeled datasets to recognize specific types of attacks, to unsupervised learning \cite{Cunningham2008}, which detects anomalies without prior knowledge of attack signatures. Another promising area involves semi-supervised learning, which combines elements of both to efficiently handle data with sparse labels \cite{zhu2005semi}. Despite the advantages these methods offer, each comes with limitations, ranging from high false positive rates in behavior-based systems to challenges of keeping supervised models up-to-date with new attack vectors.

In this work, we extend our innovative approach \cite{pitsiorlas2024latent}  to the field of network security and intrusion detection, by leveraging the capabilities of Variational Autoencoders (VAEs). VAEs, which are known for their proficiency in generating new data instances and encoding data into a compact latent space \cite{Kingma_2019}, offer a unique advantage in identifying intricate patterns indicative of cyberattacks \cite{9113298}. By employing VAEs, our methodology not only aims to detect anomalies but also to assess the reliability of unknown samples before their evaluation. This is achieved through the development of a confidence metric that provides insights into the expected accuracy of anomaly classifications, thereby addressing a critical challenge in intrusion detection research. Using the VAE's architecture, our confidence metric, based on the latent space representations, and with the core methodology, based on generating meaningful latent spaces, proved effective in enhancing the trustworthiness of our predictions.

For instance, consider a scenario where an IDS detects potential anomalies in network traffic. Traditional methods may flag these anomalies, but without a measure of confidence, the system might generate numerous false positives, leading to unnecessary alerts and wasted resources. By incorporating our confidence metric, derived from the Mahalanobis distance in the latent space, we can effectively gauge the trustworthiness of each detection. Suppose an unknown sample is considered highly trustworthy based on its proximity to known training samples in the latent space, we can reduce false positives and increase the efficiency of our IDS without any additional manual tuning or complex adjustments. This straightforward enhancement demonstrates the practical value of our approach in real-world applications.

In the current study, we adapt and refine our algorithm to address the challenges of anomaly detection in IDS. By applying our method to the NSL-KDD dataset \cite{5356528}, a benchmark dataset in network security research, we demonstrate its versatility and effectiveness across diverse data domains. This cross-disciplinary application underscores the robustness of our VAE-based technique in generating reliable confidence metrics for predictions, which is a critical aspect in both environmental studies and cybersecurity. Through this continuity of research, we not only validate the universal applicability of our method but also contribute to the advancement of ML applications in ensuring data integrity and security in networked environments. Focusing on binary classification, our work simplifies the complex landscape of network threats into a dichotomy of normal and malicious activities. This simplification allows for a more streamlined and focused approach to identifying and mitigating cyberthreats. 

The implications of our findings are important. By enhancing the reliability and accuracy of anomaly detection, our approach contributes to the development of more robust and effective IDS systems. In a broader context, this research addresses the urgent need for advanced cybersecurity measures in an increasingly interconnected world. 

\section{System Model}
We leverage the architecture of VAEs and their proficiency in generating novel data instances. A VAE is a Directed Probabilistic Graphical Model (DPGM) that has a posterior that is approximated by a neural network having an autoencoder-like architecture \cite{An2015VariationalAB}. The encoder $\text{Enc}(\cdot)$ operates on the input data $\mathbf{X}$, which encompass features such as security, temporal, and protocol among others, to encode them into a new representation projected within the latent space, i.e., $\text{Enc}(\mathbf{X}) = \mathbf{Z}$. Subsequently, the decoder $\text{Dec} (\cdot)$ utilizes this latent space representation $\mathbf{Z}$ as input to generate new data $\hat{\mathbf{X}}$, i.e., $\text{Dec}({\mathbf{Z}}) = \hat{\mathbf{X}}$. These generated data instances are accompanied by a reconstruction error $\hat{Re}$.

Mathematically, the VAE consists of an encoder network \( q_\phi(\mathbf{Z}|\mathbf{X}) \) and a decoder network \( p_\theta(\mathbf{X}|\mathbf{Z}) \), where:
\begin{itemize}
    \item $\mathbf{X}$ is the input data,
    \item $\mathbf{Z}$ is the latent representation,
    \item $\phi$ and $\theta$ are the parameters of the encoder and decoder networks, respectively.
\end{itemize}

The encoder approximates the posterior distribution \( p(\mathbf{Z}|\mathbf{X}) \) and the decoder reconstructs the input data from the latent representation:
\begin{equation}
q_\phi(\mathbf{Z}|\mathbf{X}) \approx p(\mathbf{Z}|\mathbf{X}).
\end{equation}

The objective of training a VAE is to maximize the Evidence Lower Bound (ELBO), which is given by
\begin{equation}
\mathcal{L}(\theta, \phi; \mathbf{x}) = \mathbb{E}_{q_\phi(\mathbf{Z}|\mathbf{X})}[\log p_\theta(\mathbf{X}|\mathbf{Z})] - \text{KL}(q_\phi(\mathbf{Z}|\mathbf{x}) \| p(\mathbf{Z}))
\end{equation}
where \(\text{KL}(\cdot \| \cdot)\) denotes the Kullback-Leibler (KL) divergence between the approximate posterior \( q_\phi(\mathbf{Z}|\mathbf{X}) \) and the prior \( p(\mathbf{Z}) \).

The first term of the ELBO is the reconstruction loss, which ensures that the decoder can effectively reconstruct the input data from the latent representation. The second term is the KL divergence loss, which regularizes the latent space to follow a prior distribution, typically a standard normal distribution.

\subsection{Optimization Problem}
The optimization problem we address is to find the optimal parameters \(\theta\) and \(\phi\) of the VAE that minimize the negative ELBO, i.e., 
\begin{equation}
\min_{\theta, \phi} -\mathcal{L}(\theta, \phi; \mathbf{x}).
\end{equation}

Additionally, we aim to minimize the reconstruction error for normal data while ensuring that the latent space provides a meaningful representation for detecting anomalies. This involves finding the right balance between the reconstruction loss and the KL divergence loss.

The complete objective function for training the VAE can thus be written as:
\begin{equation}
\mathcal{L}_{\text{total}} = \mathcal{L}_{\text{Re}} + \beta \mathcal{L}_{\text{KL}}
\label{eq:total_loss}
\end{equation}
where \(\beta\) is a hyperparameter that controls the trade-off between the reconstruction loss $\mathcal{L}_{\text{Re}}$ and the KL divergence loss $\mathcal{L}_{\text{KL}}$. So it captures the trade-off between accuracy and generalization.
By optimizing this objective function $\mathcal{L}_{\text{total}}$, we ensure that the VAE can effectively reconstruct normal data and provide a reliable measure for detecting anomalies based on the reconstruction error.

\subsection{Anomaly Detection}
In this work, we focus on a reconstruction-based anomaly detection method. Reconstruction-based methods try to identify anomalies by comparing the original input data to the data reconstructed by the model. The underlying assumption is that anomalies are not well-represented by the latent space learned from the normal data, and thus will have higher reconstruction errors.

Given an input sample \(\mathbf{x}\), the VAE projects it to the latent space to obtain \(\mathbf{z}\), and then reconstructs it back to \(\hat{\mathbf{x}}\). The reconstruction error \(\hat{Re}\) is computed as
\begin{equation}
\hat{Re} = \| \mathbf{x} - \hat{\mathbf{x}} \|^2.
\end{equation}

To perform anomaly detection, we set a predefined threshold \(T\) on the reconstruction error. This threshold setting is crucial as it directly influences the sensitivity and specificity of the anomaly detection. If the reconstruction error for a given sample exceeds this threshold, the sample is flagged as an anomaly:
\begin{equation}
{\hat{y}} = 
\begin{cases} 
1 & \text{if } \hat{Re} > T \\
0 & \text{otherwise}.
\end{cases}
\end{equation}

\section{Proposed Methodology}
In this study, our objective is to introduce a methodology that assesses the reliability of an unknown sample for anomaly detection before subjecting it to evaluation. Our goal is to derive a confidence metric, $\mathcal{C}$, which is informative regarding each unknown instance's $X_{un}$ expected error $\hat{e}$ and therefore the final classification as an anomaly or not, given the train set $X_{train}$ and the predictor $P(\cdot)$. The confidence metric can be expressed mathematically as
\begin{equation}
\mathcal{C}(X_{un}|X_{train}, P(\cdot)) \propto \hat{e} ~ ~.
\end{equation}

\begin{figure*}[htb]
\centering
\includegraphics[width=\textwidth]{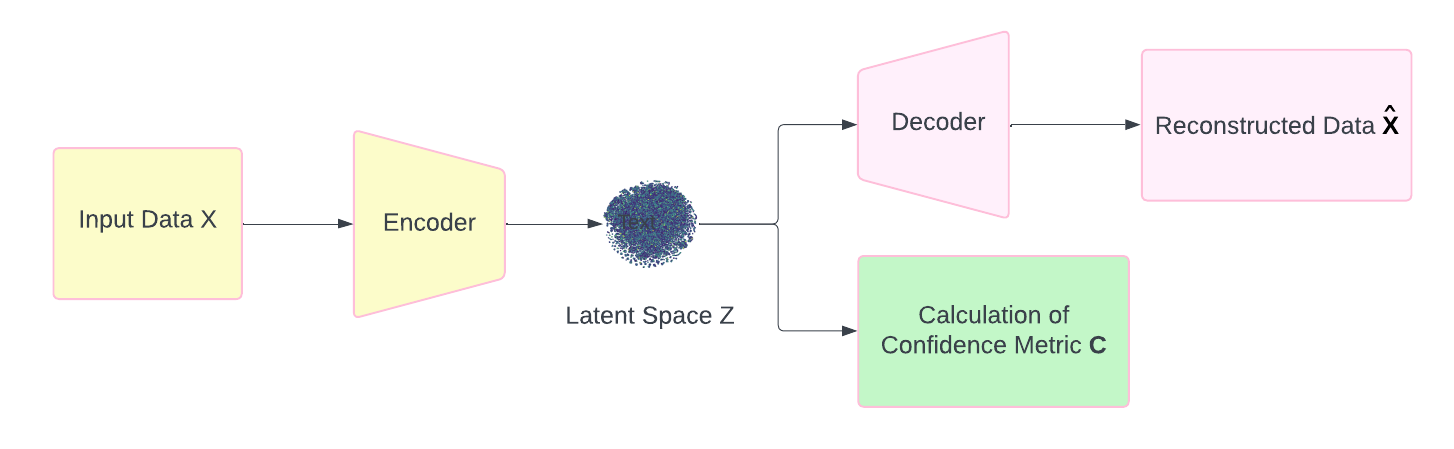}
\caption{The main components of our VAE architecture}
\label{fig:correlation_beta}
\end{figure*}

\subsection{Prediction Error Calculation and Post Hoc Classification}
The expected prediction error $\hat{e}$ is defined as the absolute value of the difference between the label $y_{true}$ and the normalized reconstruction error, $\hat{Re}$, obtained from the VAE:
\begin{equation}
\hat{e} = |y_{true} - \hat{Re}|.
\end{equation}

This means that when the instance is labeled as an anomaly, we strive for a reconstruction error that is as high as possible, and when the instance is labeled as normal, we want the reconstruction error to be as low as possible.
Our approach treats this error as a continuous variable.

\subsection{Mahalanobis distance}
Since we evaluate the reliability of unknown observations in latent space, we choose to use the Mahalanobis distance~\cite{mahalanobis}. The Mahalanobis distance offers several key advantages over the Euclidean distance.

Mathematically, the Mahalanobis distance between a point \( \mathbf{x} \) and a distribution with mean \( \boldsymbol{\mu} \) and covariance matrix \( \Sigma \) is defined as:
\begin{equation}
D_M(\mathbf{x}) = \sqrt{ (\mathbf{x} - \boldsymbol{\mu})^T \Sigma^{-1} (\mathbf{x} - \boldsymbol{\mu}) }.
\end{equation}

Here, \( \mathbf{x} \) is the latent representation of an observation, \( \boldsymbol{\mu} \) is the mean vector of the training latent representations \( Z_{\text{train}} \), and \( \Sigma \) is the covariance matrix of \( Z_{\text{train}} \).

The Mahalanobis distance accounts for the distribution of the data by incorporating the covariance structure of the latent representations. This metric effectively normalizes the distances based on the variance along each dimension, making it sensitive to the correlation between features. Consequently, it scales the distance measurements appropriately according to the actual spread of the data. 

In contrast, the Euclidean distance treats all dimensions equally, ignoring the potential variability and correlations that exist in the data, which can lead to misleading results in high-dimensional spaces where features may have different scales and variances.

By using the Mahalanobis distance, we improve our ability to discern how anomalous a new observation is relative to the training data, leading to a more robust and reliable confidence metric for anomaly detection. This method allows us to assess the reliability of each unknown sample before subjecting it to evaluation, ultimately improving the overall trustworthiness of our IDS. 

\subsection{Confidence Metric Calculation}
To calculate the confidence metric $\mathcal{C}$ using the Mahalanobis distance, we incorporate the covariance structure of the latent space. First, we train the VAE model to obtain latent representations of the training data, $Z_{\text{train}}$, and the unknown observations, $Z_{\text{un}}$. Once trained, we use the model as an evaluator for the inference data we have. During evaluation, the model generates new instances based on the features of the inference data. We consider that the unknown observations that are closer to representations of training points $Z_{train}$ are more trustworthy.

The distance $\mathcal{C}$ for the $j$-th unknown observation can be written as:
\begin{equation}
    \mathcal{C}_j = \sqrt{ (z_{\text{un}}^j - (z_{j,1}^+)^T \Sigma_{\text{train}}^{-1} (z_{\text{un}}^j - (z_{j,1}^+)) }
\end{equation}
where
\begin{itemize}
    \item $z_{\text{un}}^j$ is the latent representation of the $j$-th unknown observation,
     \item $z_{j,1}^+$ is the nearest representation to the $j$-th unknown observation,
    \item $\Sigma_{\text{train}}$ is the covariance matrix of $Z_{\text{train}}$,
    \item $\Sigma_{\text{train}}^{-1}$ is the inverse of the covariance matrix.
\end{itemize}

\subsection{Evaluation Metric}
To evaluate if the proposed confidence metric $\mathcal{C}$ is informative with the error, we measure its correlation with prediction error $\hat{e}$
\begin{equation}
r = \text{corr}(\mathcal{C}, \hat{e}) ~~.
\end{equation}

\section{Experimental Results}
\subsection{The NSL-KDD Dataset}
Among the datasets employed to benchmark IDS performance, the NSL-KDD dataset \cite{5356528} has gained prominence. As an improved and standardized version of the KDD Cup 99 dataset\cite{kddcup99}, it addresses several of its predecessor's shortcomings, offering a more reliable foundation for research in intrusion detection and network security. The NSL-KDD dataset's widespread acceptance underscores its relevance and utility in fostering advancements within this domain.

Although the NSL-KDD dataset traditionally encompasses multi-class labels, in our experiments we focus on using it as a binary classification task. Specifically, from the $41$ features that the dataset includes, regarding the attack, we consolidate all instances of network intrusions into a single class labeled as '1', while normal network traffic is labeled as '0'. This decision allows us to simplify the classification task and focus our efforts on detecting malicious activities effectively. Analyzing the dataset, we observe that the training set comprises $46.54$\% instances of network intrusions and $53.46$\% instances of normal traffic. Similarly, the testing set contains $56.92$\% instances of network intrusions and $43.08$\% instances of normal traffic, showcasing the prevalence of malicious activities in both training and testing data. 

We pay close attention to the NSL-KDD dataset's details, especially how it organizes network information into different features. These features include basic details about network connections and more advanced information that helps us understand the flow of network traffic better. By looking at these features, we can tell apart normal network activities from those that might be harmful.

When we use the NSL-KDD dataset to separate normal activities from possible threats, it is not just about the features themselves but also about how these features interact with each other. This interaction helps us spot suspicious patterns. Before even starting our experiments, we prepare the dataset carefully, i.e., we make sure that all data are on a similar scale so that our analysis is accurate. 

\subsection{VAE Architecture and Training Curriculum}
The VAE architecture utilized in our study consists of two main components: an encoder and a decoder, each structured with five fully connected layers. The encoder is responsible for transforming the input data into a lower-dimensional latent space representation. It begins with an input layer that matches the dimensionality of the input data, followed by a sequence of linear layers with respective dimensions of $512$, $384$, $256$, and $128$ neurons, each activated by Rectified Linear Unit (ReLU) functions. The final layer in the encoder outputs two sets of values: one for the mean and one for the log variance of the latent space distribution, each with a dimension equal to twice the latent dimension size to accommodate the mean and log variance parameters.

The decoder, on the other hand, reconstructs the original input data from the latent space representation. It starts with a linear layer that takes the latent dimension as input and expands it to 128 neurons, followed by layers of $256$, $384$, and $512$ neurons, each activated by rectified linear unit (ReLU) functions. The final layer in the decoder maps the $512$-dimensional representation back to the original input dimension.

Given the unsupervised nature of our task, the data consists only of features without explicit labels. During the training phase, we iterate over $30$ epochs, using an initial learning rate of $0.001$, dynamically adjusted by a StepLR scheduler. The optimization process employs the Adam Optimizer. The VAE's loss function combines the Mean Squared Error (MSE) loss, which quantifies the difference between the reconstructed output and the original input, with the KL divergence loss.

\begin{figure}[htb]
\centering
\begin{minipage}[b]{.49\linewidth}
\centering
\includegraphics[width=4cm]{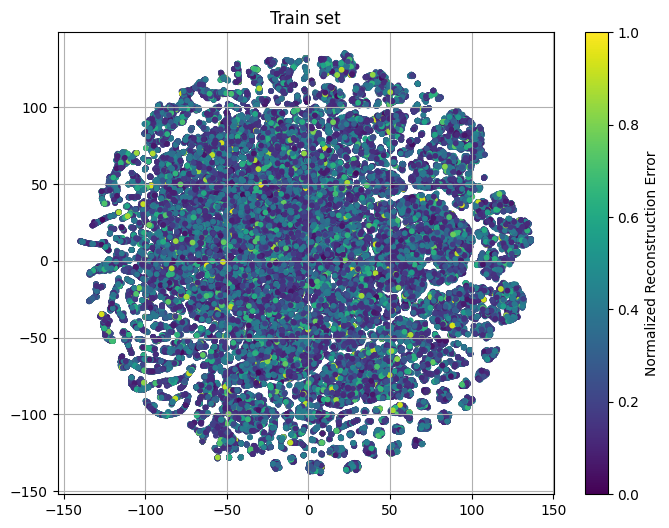}
\subcaption{Train Set}
\label{fig:train}
\end{minipage}
\hfill
\begin{minipage}[b]{.49\linewidth}
\centering
\includegraphics[width=4cm]{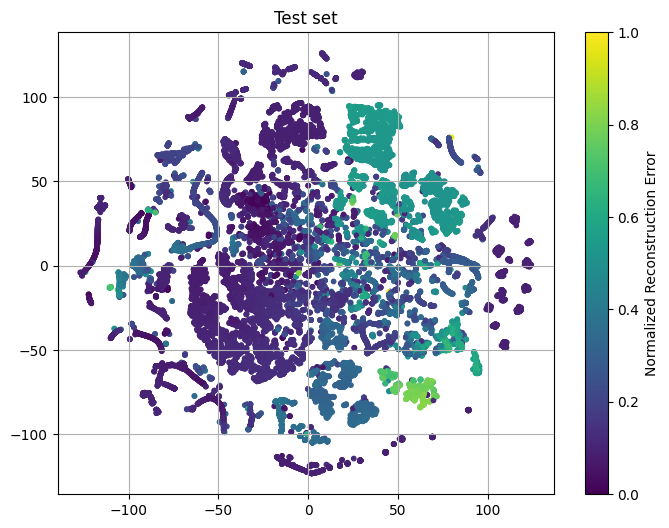}
\subcaption{Test Set}
\label{fig:test}
\end{minipage}
\caption{T-SNE Visualization of the latent space for both train and test sets.}
\label{fig:latent}
\end{figure}

Figure \ref{fig:latent} illustrates the latent space projections of both train and test data, once the predictor is trained with the available $X_{train}$. The color coding in the figure helps depict the prediction error of each observation, providing a visual representation of how well the VAE model manages to cluster and differentiate between normal and anomalous instances.

\subsection{Adjusting Model Parameters}
Two main parameters affect the performance of our confidence metric: (i) the dimension of the latent space, and (ii) the $\beta$ parameter that controls the trade-off between accuracy and generalization. 
The selection of those optimal parameters can be seen as a multivariate problem and can be approached with alternating optimization, by optimizing iteratively one parameter at a time.

\subsubsection{Latent Space Dimension}
In determining the appropriate dimension for the latent space, as we can observe in Table \ref{tab:latent_space_correlations}, we carefully consider the trade-off between computational efficiency and the efficacy of our model's performance metrics. Following a series of experiments, we conclude that setting the latent space dimension to $20$ and the default $\beta$= 1 yields near-optimal results.  
\begin{table}[htbp]
\caption{Latent Space Dimension Correlations}
\begin{center}
\begin{tabular}{|c|c|c|c|}
\hline
\textbf{LS Dimension} & \textbf{General Corr} & \textbf{Corr FN} & \textbf{Corr FP} \\
\hline
5 &26\% & 3\% & 50\% \\
\hline
10 & 39\% &2\% & 17\% \\
\hline
20 & \textbf{42}\% & 6\% & 55\% \\
\hline
30 &41\% & 12\% & 52\% \\
\hline
\end{tabular}
\end{center}
\label{tab:latent_space_correlations}
\end{table}

The results of our correlation analysis underscore the significant potential of employing the Mahalanobis distance in the latent space for improving the reliability of anomaly detection in IDSs. The general correlation of $r$ = 42\% between the latent space distance and the proposed prediction error highlights the effectiveness of the latent space representation in capturing substantial variability within the data. This indicates that the VAE model is proficient at distinguishing between normal and anomalous instances. The high correlation associated with false positives suggests that the latent space excels in capturing features of normal network traffic, which reduces the number of benign instances misclassified as attacks. This is further evidenced by the model's high precision score of $0.90$, demonstrating a strong ability to correctly identify true positive instances of network intrusions.

\subsubsection{KL weight $\beta$}
In our experiments, we adjust the KL divergence weight, denoted as 
$\beta$ in equation (\ref{eq:total_loss}), to observe its impact on the performance of the VAE. The KL weight controls the balance between the reconstruction loss and the regularization term in the VAE's loss function. By varying  $\beta$, we aim to find the optimal trade-off that enhances the model's ability to detect anomalies.

\begin{figure}[htb]
\centering
\includegraphics[width=9cm]{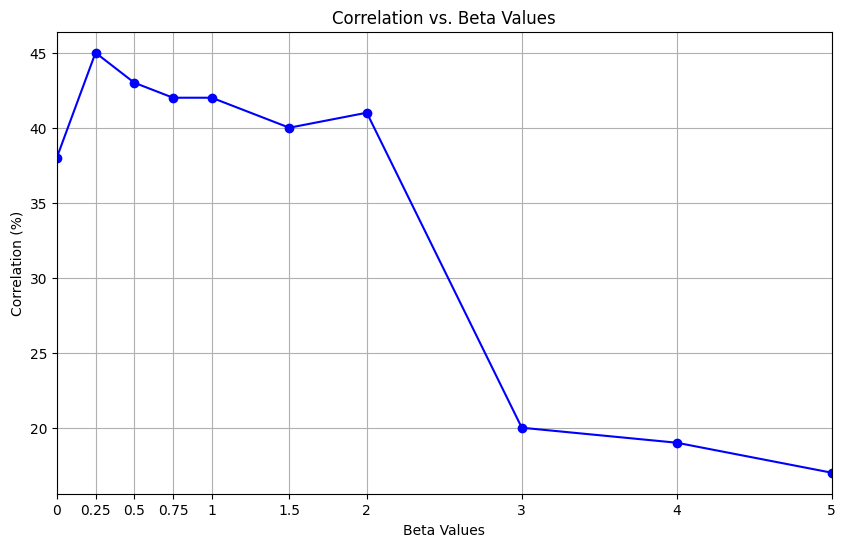}
\caption{Weights for the KL loss and correlation}
\label{fig:correlation_beta}
\end{figure}

As we can observe in Figure \ref{fig:correlation_beta}, the optimal value of $\beta$ is $0.25$. The results indicate that this value yields the highest correlation for our metric of \textbf{45\%}.  As $\beta$ goes higher we can see that the correlation drops significantly, indicating that too much regularization can hinder the model's effectiveness in capturing meaningful patterns in the latent space for anomaly detection.

This choice strikes as well a balance between computational complexity and the ability of the model to effectively capture the underlying structure of the data, resulting in favorable outcomes across our proposed evaluation metric.

\subsection{Optimal Threshold and Anomaly Predictions}
To determine the optimal threshold for the reconstruction error and subsequent classification, we conduct an analysis aiming to maximize the F1 score on the training data by having the dimension of the latent space equal to $20$ and $\beta$ = 0.25. Through this process, we identify an optimal threshold T value of approximately $T$ = $0.07$. We can observe the evaluation metrics below.
\begin{itemize}
\item Precision: 0.9093
\item Recall: 0.6444
\item F1-score: 0.7543
\item Accuracy: 0.7610.
\end{itemize}
Despite its seemingly small magnitude, this threshold effectively balances precision and recall, resulting in robust classification performance on the training dataset.

It is also interesting to examine the results of the confusion matrix:

\begin{figure}[htb]
\centering
\includegraphics[width=8cm]{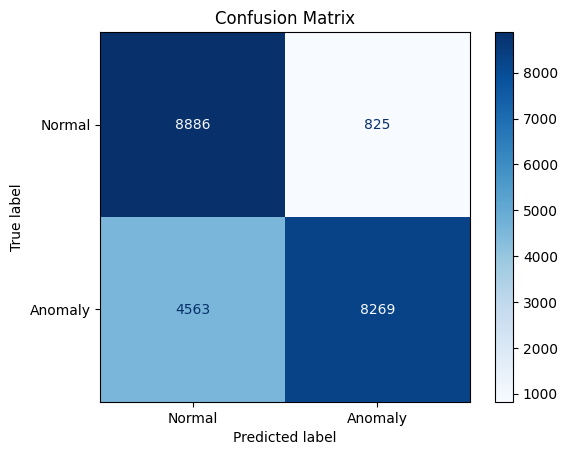}
\caption{Confusion Matrix}
\label{fig:correlation_beta}
\end{figure}

\begin{table}[htbp]
\caption{Correlations of Latent Space Dimension with Different Error Types}
\begin{center}
\begin{tabular}{|c|p{1cm}|p{1cm}|p{1cm}|p{1cm}|p{1cm}|}
\hline
\textbf{LS Dim.} & \textbf{Corr FP} & \textbf{Corr FN} & \textbf{Corr TP} & \textbf{Corr TN} & \textbf{Gen. Corr} \\
\hline
20 & 57\% & 7\% & 15\% & 19\% & \textbf{45\%} \\
\hline
\end{tabular}
\end{center}
\label{tab:latent_space_detailed_correlations}
\end{table}



Table \ref{tab:latent_space_detailed_correlations}
presents the correlation analysis between the confidence metric and various error types (False Positives, False Negatives, True Positives, and True Negatives) for the optimal latent space dimension of $20$. The correlation values indicate how well the confidence metric, derived from the latent space distance, aligns with the actual classification errors in the anomaly detection task.

The correlation for False Positives (FP) is notably high at 57\%, suggesting that the confidence metric is particularly effective in identifying normal instances that are incorrectly classified as anomalies. This high correlation indicates that the metric can help reduce false positives by providing a reliable measure of how similar an instance is to the training data.

In contrast, the correlation for False Negatives (FN) is relatively low at 
7\%, implying that the confidence metric is less effective in detecting anomalous instances that are incorrectly classified as normal. This indicates a potential area for improvement in the model's ability to correctly identify true anomalies.

The correlations for True Positives (TP) and True Negatives (TN) are 
15\% and 19\%, respectively, showing a moderate relationship between the confidence metric and these correct classifications. The general correlation across all error types is 
45\%, reflecting the overall effectiveness of the confidence metric in capturing the variability in the data for anomaly detection.

These results underscore the importance of the confidence metric in improving detection accuracy, particularly in reducing false positives, and highlight the need for further refinement to enhance the detection of true anomalies.

\subsection{Experiment with Choquet–Mahalanobis distance}
In addition to our primary approach, we conduct experiments comparing the effectiveness of the Mahalanobis and the Choquet Integral Operator \cite{TORRA201256}. The Choquet integral is a sophisticated aggregation operator that can capture interactions between features in a more refined manner. This method is particularly advantageous in scenarios where feature interactions are complex and not easily captured by traditional distance measures. For this experiment, we implement the Choquet integral to aggregate the distances between the latent representations of the training and test sets.

Our findings indicate that the Choquet–Mahalanobis distance can achieve a general correlation of  45\% between the confidence metric and the prediction error. 
However, it is important to note that the Choquet integral, while effective, proved to be significantly more computationally expensive compared to the Mahalanobis distance. The increased computational cost arises from the complexity involved in computing the Choquet integral, which requires more intensive calculations to capture the interactions between features.

\subsection{Evaluation of the Confidence Metric}
To evaluate the effectiveness of our confidence metric in assessing anomaly detection, we conduct a comprehensive analysis across multiple dimensions. In addition to assessing the distance in the Latent Space (LS), we also calculated distances in the Feature Space (FS).

\subsubsection{Comparison of different distances}
In our analysis, we evaluate the effectiveness of three different distance metrics: Mahalanobis, Euclidean, and Cosine. The Mahalanobis distance, which as stated before, takes into account the covariance structure of the latent space, demonstrates the highest correlation of 45\% with the error metric. This suggests that the Mahalanobis distance is particularly effective in capturing the intricate patterns in the latent space that are indicative of anomalies. The Euclidean distance, a more straightforward metric, shows a correlation of 38\%. While it is computationally less complex, it does not account for the underlying data distribution, which may lead to less accurate anomaly detection compared to Mahalanobis distance. The Cosine distance, which measures the cosine of the angle between two vectors, exhibits a correlation of 32\%. This metric is often used for high-dimensional spaces and can be useful when the magnitude of the vectors is less important than their direction. However, in our context, it proves to be less effective than both Mahalanobis and Euclidean distances. Overall, the Mahalanobis distance provided the best performance in terms of correlation with the error metric, indicating its superiority for our specific use case in IDS anomaly detection. 

\begin{table}[h!]
\centering
\caption{Comparison of different distances} \label{tab:comp_dist}
\begin{tabular}{llll}
\hline
 & \textbf{Mahalanobis} & \textbf{Euclidean} & \textbf{Cosine} \\ \hline
\textbf{Correlation} & \textbf{45\%} & 38\% & 32\% \\ \hline
\end{tabular}
\end{table}

\subsubsection{Computational Complexities of Mahalanobis Distance in LS vs. FS}
When implementing Mahalanobis distance, the computational complexity for the LS and the FS can vary significantly. This disparity primarily arises due to the dimensional differences and the covariance matrix computation. In latent space, where the dimensionality is typically much lower due to the compressed representations learned by the VAE, the computation of the covariance matrix and its inverse is computationally less demanding. Consequently, calculating Mahalanobis distances in this space tends to be faster, as evidenced by the observed execution time of approximately 8 minutes. In contrast, the feature space often retains the high-dimensional nature of the original data. This increases the computational burden associated with estimating the covariance matrix, regularizing it, and performing matrix inversion. Additionally, the high dimensionality amplifies the cost of pairwise distance calculations, leading to a noticeable increase in processing time, with the feature space calculations taking around 37 minutes. This significant difference highlights the practical advantage of using latent space representations for distance calculations, which not only provide meaningful lower-dimensional embeddings but also enhance computational efficiency.

\section{Discussion}
This research extends an innovative use of Variational Autoencoders to the critical field of network security, demonstrating the VAE's capability not only to distill complex network traffic data into a meaningful latent space but also to enhance the reliability of anomaly detection in Intrusion Detection Systems. The introduction of a reliable confidence metric derived from latent space representations marks a significant leap forward in our methodology, offering a refined approach to evaluating the trustworthiness of IDS predictions.

The notable correlation observed in the latent space underlines the VAE model's profound ability to capture and interpret the complex patterns indicative of cyberthreats. This correlation is not merely a numerical assessment but a reflection of the intrinsic data structure, which significantly contributes to the confidence in anomaly detection. The higher correlation values in the latent space, compared to those in the feature space, underscore the latent space's critical role in identifying and distinguishing between normal and malicious network activities.

Our findings demonstrate the latent space's potential beyond dimensionality reduction, establishing it as a pivotal element for ensuring prediction reliability in cybersecurity applications. This methodology bridges the gap between raw network data and actionable insights, enhancing the IDS capability to safeguard against sophisticated cyberattacks while fostering trust in ML-based security solutions.

Comparing our approach to traditional methods of anomaly detection, which often rely on either predefined rules or supervised learning models, our VAE-based methodology offers a dynamic solution capable of adapting to new and evolving threats. Unlike many existing models that struggle with high false positive rates or require extensive labeled datasets, our method effectively utilizes unsupervised learning to identify intricate patterns indicative of cyberattacks, as supported by the correlation metrics presented in our results.

Future research should aim to further validate and refine the proposed methodology across a broader range of datasets and network environments. Investigating the scalability of this approach to handle multi-class classification tasks and exploring ways to reduce computational overhead are critical next steps. Additionally, integrating our confidence metric with real-time monitoring systems could offer new insights into its practical applicability and effectiveness in operational settings.

\section*{Acknowledgments}
The work of I. Pitsiorlas and M. Kountouris has been supported by the EC through the Horizon Europe/JU SNS project, ROBUST-6G (Grant Agreement no. 101139068).

\bibliographystyle{IEEEbib}
\bibliography{refs}

\end{document}